\documentclass[nofootinbib,twocolumn,showpacs,aps,preprintnumbers,amsmath,amssymb,superscriptaddress]{revtex4-1}
 \usepackage{graphicx,subfigure,multirow,epsfig}
 
\usepackage{dcolumn}
\usepackage{bm}
\usepackage{color}
 
\makeatletter

\newcommand{\fmslash}[2][0mu]{%
  \mathchoice
    {\fmsl@sh\displaystyle{#1}{#2}}%
    {\fmsl@sh\textstyle{#1}{#2}}%
    {\fmsl@sh\scriptstyle{#1}{#2}}%
    {\fmsl@sh\scriptscriptstyle{#1}{#2}}}
\newcommand{\fmsl@sh}[3]{%
  \m@th\ooalign{$\hfil#1\mkern#2/\hfil$\crcr$#1#3$}}
\makeatother

 \newcommand{\lsim}{{\;\raise0.3ex\hbox{$<$\kern-0.75em\raise-1.1ex\hbox{$\sim$}}\;}}
\newcommand{\gsim}{{\;\raise0.3ex\hbox{$>$\kern-0.75em\raise-1.1ex\hbox{$\sim$}}\;}}

\newcommand{\beq}{\begin{equation}}
\newcommand{\eeq}{\end{equation}}
\newcommand{\bea}{\begin{eqnarray}}
\newcommand{\eea}{\end{eqnarray}}
\mathchardef\minus="002D
  
\newcommand{\met}{{\fmslash E_T}} 
\allowdisplaybreaks

\begin{document}
\title{Probing resonance decays to two visible and multiple invisible
    particles}
    
\author{Won Sang Cho}
\affiliation{Physics Department, University of Florida, Gainesville, FL 32611, USA}
\author{Doojin Kim}
\affiliation{
Department of Physics, University of Maryland, College Park, MD 20742, USA}
\author{Konstantin T.~Matchev}
\affiliation{Physics Department, University of Florida, Gainesville, FL 32611, USA}
\author{Myeonghun Park}
\affiliation{CERN, Theory Division, CH-1211 Geneva 23, Switzerland}
\preprint{CERN-PH-TH/2012-160}
\preprint{UMD-PP-012-006}
\date{March 5, 2014}

\begin{abstract}
We consider the decay of a generic resonance to two visible particles and any number of invisible particles.
We show that the shape of the invariant mass distribution of the two visible particles is sensitive to both the
mass spectrum of the new particles, as well as the decay topology. We provide the analytical formulas describing the
invariant mass shapes for the nine simplest topologies (with up to two invisible particles in the final state). 
Any such distribution can be simply categorized by its endpoint, peak location and curvature, which are
typically sufficient to discriminate among the competing topologies.
In each case, we list the effective mass parameters which can be measured by experiment.
In certain cases, the invariant mass shape is sufficient
to completely determine the new particle mass spectrum, including the overall mass scale.
\end{abstract}





\maketitle

The dark matter problem and the mystery of the feeble neutrinos greatly motivate the ongoing LHC
searches for new physics in channels with missing energy. 
Alas, at hadron colliders like the LHC, deciphering events
with invisible particles in the final state is notoriously difficult.

\begin{figure}[t]
\includegraphics[width=3.8cm]{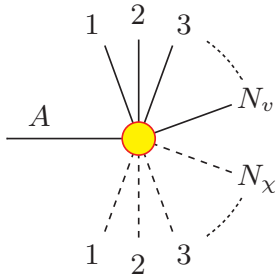}  
\caption{\label{fig:diagram} The generic decay topology under consideration.}
\end{figure}

The problem is schematically illustrated in Fig.~\ref{fig:diagram}, 
which depicts the generic decay of some new heavy resonance $A$ 
into $N_{v}$ visible particles $v_i$ and $N_{\chi}$ ``invisible" particles 
$\chi_i$ (neutrinos or dark matter candidates) which leave no trace in the detector. 
A priori we have no way of knowing the underlying physics behind Fig.~\ref{fig:diagram},
and thus we are missing the answers to some very basic questions:
1) How many invisible particles are in the final state?
2) What are their masses?
3) What is the exact topology (i.e. Feynman diagram) of Fig.~\ref{fig:diagram}:
are there any intermediate resonances, and if so, what are their masses?

Historically, the topic of mass measurements
has attracted the most attention in the literature (for a review, see \cite{Barr:2010zj}).
Unfortunately, virtually all proposed methods suffer from two
drawbacks. First, one must typically assume the correct decay topology for Fig.~\ref{fig:diagram},
{\it including} the correct number $N_{\chi}$ of invisible particles. 
If this guess is incorrect, the method does not apply. 
This motivates us to address the issue of the correct decay topology and 
number of invisibles concurrently with (perhaps even prior to) the 
more traditional question of mass measurements \cite{Bai:2010hd}.
Second, most methods for mass measurements utilize kinematic endpoints, 
where the available statistics can be rather poor 
(in the sense that the most populated bins are rarely near the kinematic endpoint).
Here we shall instead concentrate on the region near the {\it peak} rather than 
the endpoint of the kinematic distribution. (The exact shape {\it in the vicinity of} the endpoint 
does contain information about $N_{\chi}$ \cite{BK,Agashe:2010tu},
but is difficult to measure precisely in the presence of backgrounds 
and detector effects.) Our main result will be the derivation of the analytical formulas 
necessary to analyze the full shape of the invariant mass distributions of the visible particles 
in Fig.~\ref{fig:diagram}, including the location of the peak.
We shall then demonstrate how those results can be used to 
determine: 1) the number of missing particles; 2) their masses; 
and 3) the associated event topology. 

\begin{table}[t]
\caption{\label{tab:counting}
The number of inequivalent event topologies as a function of $1\le N_{v} \le 4$ and $1\le N_{\chi} \le 5$. }
\begin{ruledtabular}
\begin{tabular}{c | c  c  c  c  c }
\multicolumn{1}{c|}{}                       & \multicolumn{5}{c}{$N_{\chi}$}   \\ 
\hline
$N_{v}$   &    1       &  2     &     3   &      4   &   5        \\   \hline\hline
1             &    1       &  2     &     4   &      8   &    16        \\
2             &    2       &  7    &     20  &    55   &  142   \\
3             &    4       &  20  &    78   &  270    & 860           \\
4             &    8       &  55  &    270 &  1138    & 4294   \\     
 \end{tabular}
\end{ruledtabular}
\end{table}

\begin{figure}[t]
\includegraphics[width=8.0cm]{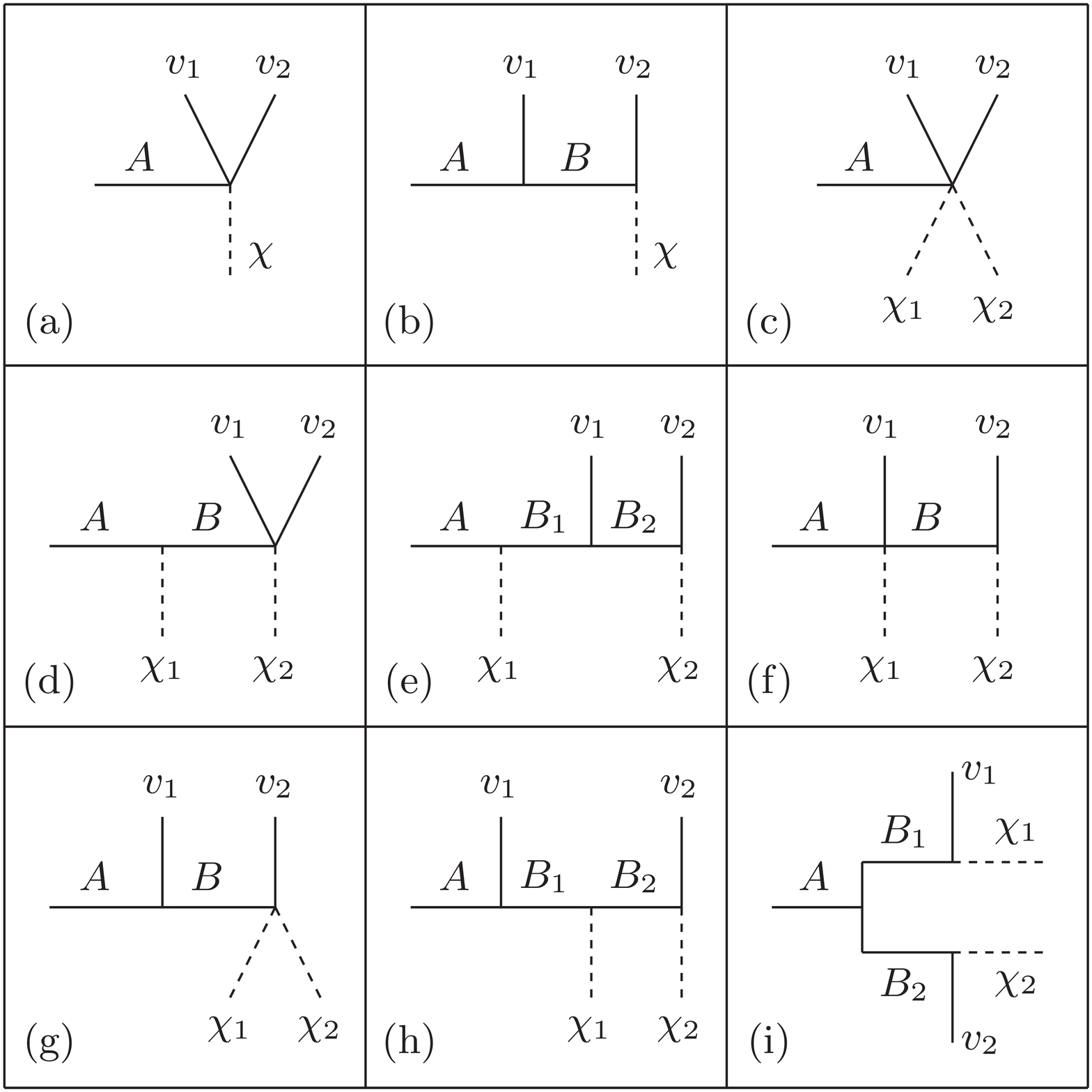}  
\caption{\label{fig:topologies} The nine $N_{v}=2$ topologies with $N_{\chi}\le2$.}
\end{figure}

Our setup is as follows. We consider the generic decay 
from Fig.~\ref{fig:diagram} without any prior assumptions about the 
decay topology or the number of invisibles. As seen in Table~\ref{tab:counting},
the number of inequivalent decay topologies proliferates very quickly as we increase 
the number of particles in the final state. Let us begin with the simplest and most challenging case
of $N_{v}=2$, postponing $N_{v}>2$ to a future study \cite{CKMP}. 
According to Table~\ref{tab:counting}, there are 2 topologies with $N_{\chi}=1$, 
shown in Fig.~\ref{fig:topologies}(a,b), and 7 topologies with $N_{\chi}=2$,
shown in Fig.~\ref{fig:topologies}(c-i). Our main goal is to analyze and contrast
the $v_1v_2$ invariant mass distribution\footnote{We note that the 
resonance $A$ is in general allowed to be produced fully inclusively, 
with an arbitrary number of {\it additional} visible or invisible particles
recoiling against $A$ in the event. This precludes us from using the
$\met$ measurement, since it will be corrupted by the invisible recoils,
which leaves us with $m_{v_1v_2}$ as the only viable observable to study.
The related combinatorial problem of partitioning the visibles in the event 
was addressed in \cite{Bai:2010hd,Blanke:2010cm}.} in each of those nine cases. 

The differential distribution of the invariant mass $m\equiv m_{v_1v_2}$
will be described by an analytical formula
\beq
\frac{dN}{dm}\equiv f(m;M_A,M_{B_i},M_{\chi_j}),
\label{eq:dNdm}
\eeq
which is only a function of the unknown masses\footnote{Note that some 
of the event topologies in Fig.~\ref{fig:topologies} involve effective higher-dimensional 
interactions \cite{Birkedal:2005cm,threebody}, which we assume to be point-like, 
otherwise effects of their mediators can be seen in other processes at the LHC.}.
Given the general formula (\ref{eq:dNdm}) for $f(m)$, 
we can easily obtain its kinematic endpoint 
\beq
E \equiv \max \left\{ m\right\}
\label{eq:E}
\eeq
and the location $P$ of the peak of the $f(m)$ distribution
\beq
f(m=P)\equiv \max \left\{ f(m)\right\}.
\eeq
Let us also define the dimensionless derivative ratios
\beq
R_n\equiv - \left( \frac{m^n}{f(m)}\frac{d^{n} f(m)}{dm^n}\right)_{m=P}.
\label{eq:Rn}
\eeq
By definition, $R_1=0$, as long as $f(m)$ is continuously differentiable at  
$m=P$, while $R_2$ parameterizes the curvature of $f(m)$ at $m=P$.

The parameters $E$, $P$ and $R_n$ are in principle all experimentally measurable from the
distribution (\ref{eq:dNdm}). Traditional studies \cite{Hinchliffe:1996iu}
have always concentrated on measuring just the endpoint $E$, failing
to utilize all of the available information encoded in the 
distribution $f(m)$. The endpoint approach gives a single measurement (\ref{eq:E}), 
which is clearly insufficient to determine the full spectrum of resonances involved 
in the decay chain of Fig.~\ref{fig:diagram}. Here we propose to invoke the full shape
(\ref{eq:dNdm}) in the analysis \cite{Birkedal:2005cm}. We envision that in practice this will be done 
by performing unbinned maximum-likelihood fits of (\ref{eq:dNdm}) to the observed data. 
In order to illustrate the power of the method here, it is sufficient to consider just
the additional individual measurements of $P$ and $R_2$.
Since they are obtained from the most populated bins near the peak,
we can expect that they will be rather well measured. 
More importantly, the additional information about
$P$ and $R_2$ might be sufficient to completely determine the mass spectrum
(see eqs.~(\ref{MAafrompeak},\ref{Mchiafrompeak}) below).
But first we need
to present our results for (\ref{eq:dNdm}-\ref{eq:Rn}) in each of the nine cases in Fig.~\ref{fig:topologies}.

{\it The topology of Fig.~\ref{fig:topologies}(a).}  For a three body decay to massless visible particles, one has
\beq
f(m;M_A,M_\chi) \sim m\, \lambda^{1/2}\left(m^2,M_A^2,M_\chi^2\right),
\label{eq:fa}
\eeq
where
\beq
\lambda(x,y,z)\equiv x^2+y^2+z^2-2xy-2yz-2xz.
\eeq
In this case
\bea
E &=& M_A-M_\chi,   
\label{eq:Ea}
\\
P &=& \left[2M_AM_\chi\left(2-\sqrt{1+3\alpha^2}\right)/(3\alpha)\right]^{1/2},
\label{eq:Pa}\\
R_2 &=& 6\left[1+\left(1+3\alpha^2\right)^{\minus 1/2}\right]^{\minus 1},
\label{eq:R2a}
\eea
where
\beq
\alpha \equiv 2M_AM_\chi/(M_A^2+M_\chi^2).
\eeq
Contrary to popular belief, one can now solve for both masses $M_A$ and $M_\chi$, given 
two of the three measurements (\ref{eq:Ea}-\ref{eq:R2a}). 
For example, using the peak location $P$ and the endpoint $E$, we find
\bea
M_A &=& \frac{E}{2}\left( \frac{P}{E} \sqrt{\frac{2-3(P/E)^2}{1-2(P/E)^2}}+1\right),  \label{MAa}\\
M_\chi &=& \frac{E}{2}\left( \frac{P}{E} \sqrt{\frac{2-3(P/E)^2}{1-2(P/E)^2}}-1\right). \label{Mchia}
\eea
Eqs.~(\ref{MAa},\ref{Mchia}) offer a new method of determining {\it both} $M_A$ and $M_\chi$,
which is a simpler alternative to the $M_{T2}$ kink method of \cite{Cho:2007qv}, 
since here we do not rely on the $\met$ measurement at all, 
and do not require to reconstruct the decay chain on the other side of the event.

In fact, one does not even need an endpoint measurement, since
the peak location $P$ and the curvature $R_2$ are sufficient for this purpose:
\bea
M_A &=& \frac{P}{\sqrt{2}}\left( \frac{6-R_2}{4-R_2} +\sqrt{\frac{12-R_2}{4-R_2}}\right)^{1/2},  \label{MAafrompeak}\\
M_\chi &=& \frac{P}{\sqrt{2}} \left( \frac{6-R_2}{4-R_2} -\sqrt{\frac{12-R_2}{4-R_2}}\right)^{1/2}. \label{Mchiafrompeak}
\eea
Note that, in analogy to the matrix element method \cite{Alwall:2009sv},
eqs.~(\ref{MAafrompeak},\ref{Mchiafrompeak}) 
are capable of determining the complete mass spectrum 
in a short SUSY-like decay chain, without relying on any kinematic endpoint measurements.

\begin{figure}[t]
\includegraphics[width=4.28cm]{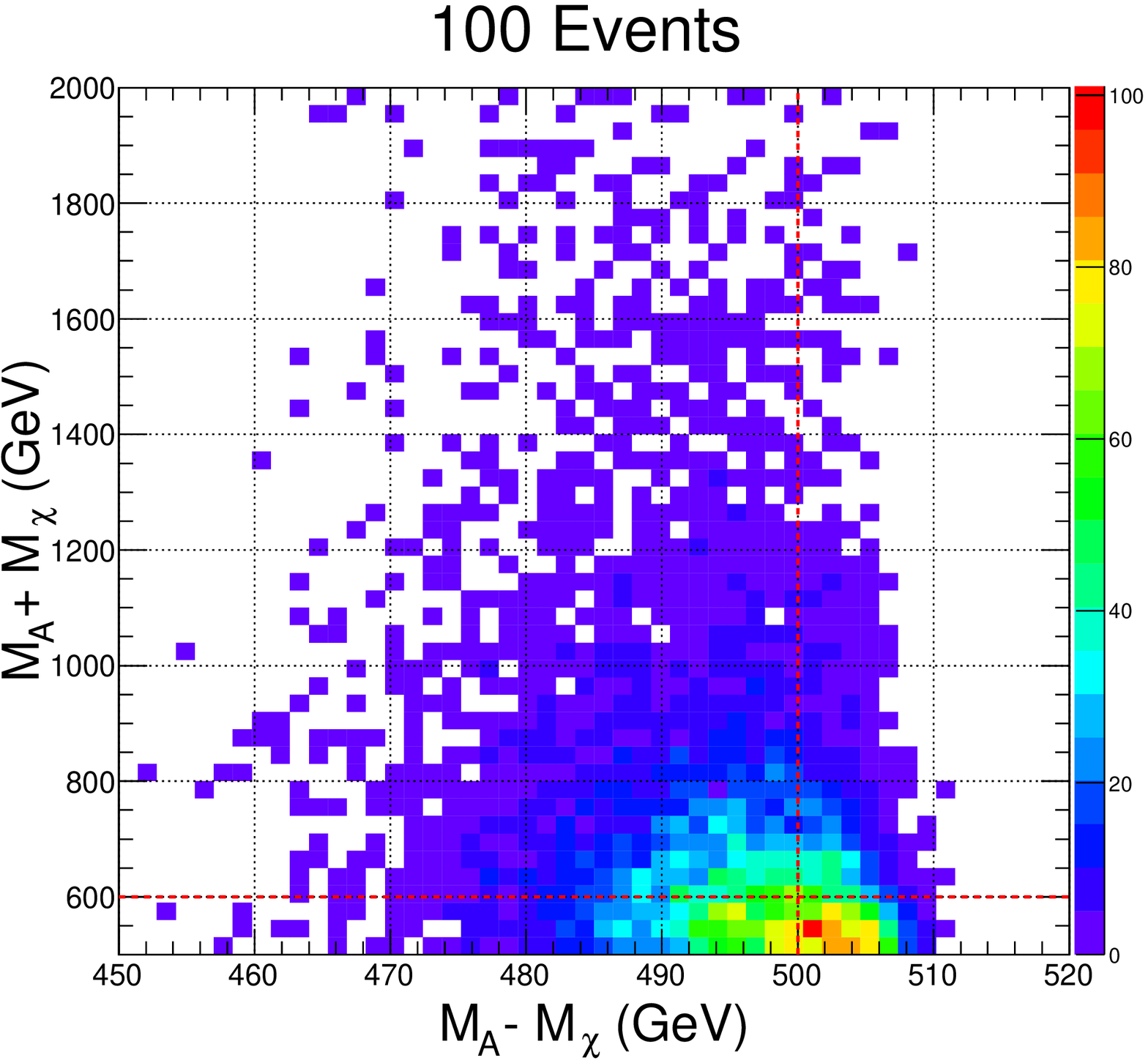}
\includegraphics[width=4.27cm]{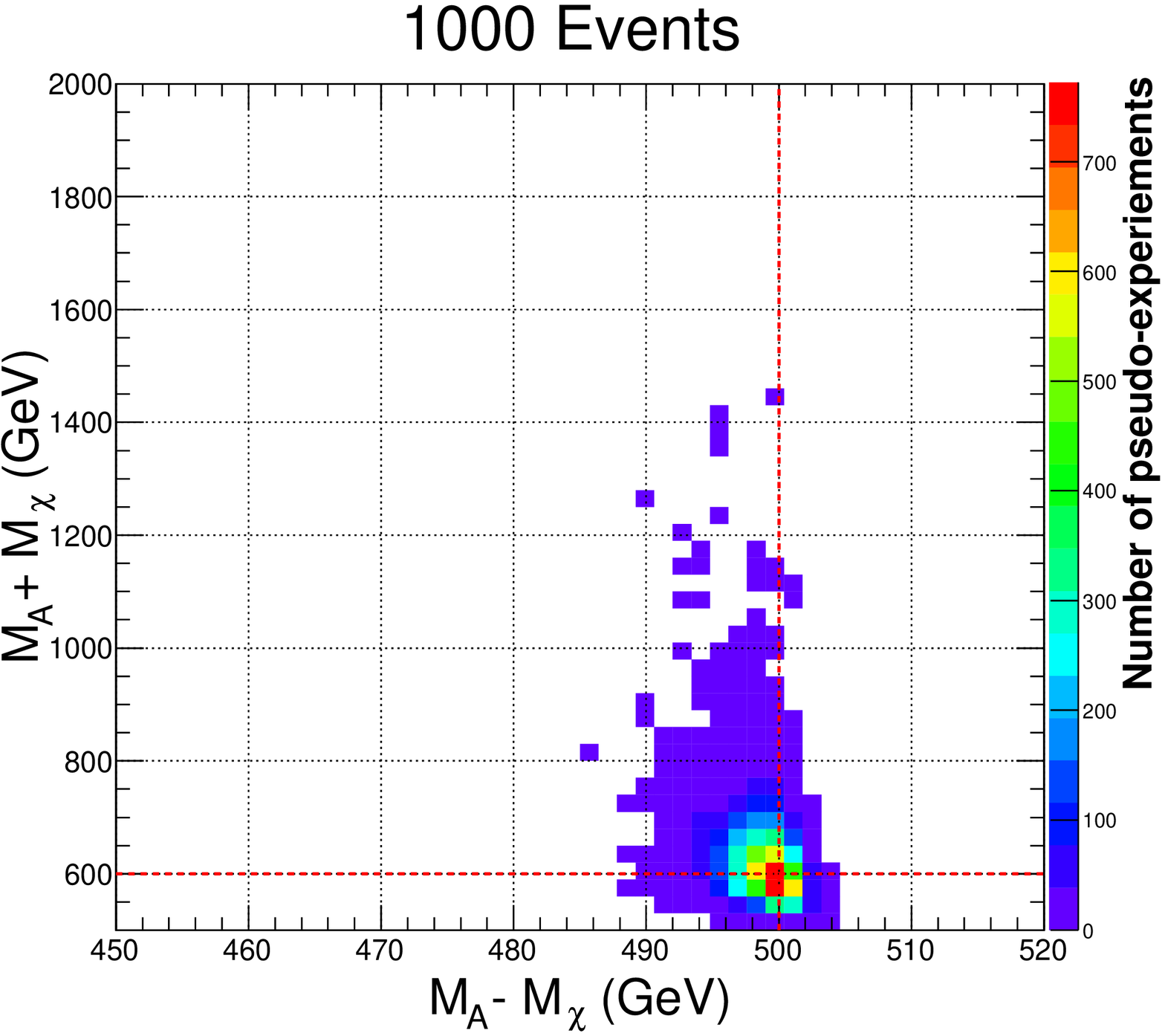}
\caption{Distribution of $M_A$ and $M_\chi$
found by a maximum-likelihood fit to eq.~(\ref{eq:fa})
in 10,000 pseudo-experiments with 100 signal events (left) or
1000 signal events (right). The input study point has $M_A=550$ GeV 
and $M_\chi = 50$ GeV. \label{fig:exp} }
\end{figure}

In order to get a rough idea of the precision of these mass determinations,
in Fig.~\ref{fig:exp} on the left (right) we show the results from 10,000 
pseudo-experiments with 100 (1000) signal events each. In each 
pseudo-experiment, the two masses $M_A$ and $M_\chi$
are extracted from a maximum-likelihood fit of the simulated data 
to the full distribution (\ref{eq:fa}). Fig.~\ref{fig:exp} shows that, as expected,
the mass difference is measured quite well, at the level of $\sim 1\%$
with just 100 events. At the same time, the mass sum (or equivalently, the
absolute mass scale) is also being determined, albeit less precisely:
at the level of $\sim 30\%$ ($\sim 10\%$) with 
100 (1000) events.

{\it The topology of Fig.~\ref{fig:topologies}(b).} Here one obtains the 
celebrated triangular shape
\beq
f(m) \sim m ,
\label{eq:fb}
\eeq
\bea
E &=& P =  \sqrt{(M_A^2-M_{B}^2)(1-M_\chi^2/M_{B}^2)} \, , 
\label{eq:Eb}
\\
R_2 &=& \infty.
\label{eq:R2b}
\eea
 
Unfortunately, the masses enter the shape (\ref{eq:fb}) only through the
combination (\ref{eq:Eb}), which is the single effective mass parameter accessible experimentally.

{\it The topology of Fig.~\ref{fig:topologies}(c).} The shape is more conveniently given in integral form,
which is easy to code up:
\bea 
 f(m) &\sim &m\int^{\scriptscriptstyle{(M_A-m)^2}}_{\scriptscriptstyle{(M_{\chi_1}+M_{\chi_2})^2}}\frac{ds}{s}
\sqrt{\lambda({\scriptstyle M_A^2,m^2,s})\lambda({\scriptstyle s,M_{\chi_1}^2,M_{\chi_2}^2})} , ~~~~~
\label{eq:fc} \\
E &=& M_A-M_{\chi_1}-M_{\chi_2} .
\label{eq:Ec}
\eea
The explicit formulas for $P$ and (\ref{eq:fc}) will be shown in \cite{CKMP}.
The important point is that in principle all three masses $M_A$, $M_{\chi_1}$ {\it and} $M_{\chi_2}$
can be simultaneously determined from a fit of eq.~(\ref{eq:fc}) to the data, just like in 
Fig.~\ref{fig:exp} \cite{CKMP}.

\begin{figure}[t]
\includegraphics[width=6.5cm]{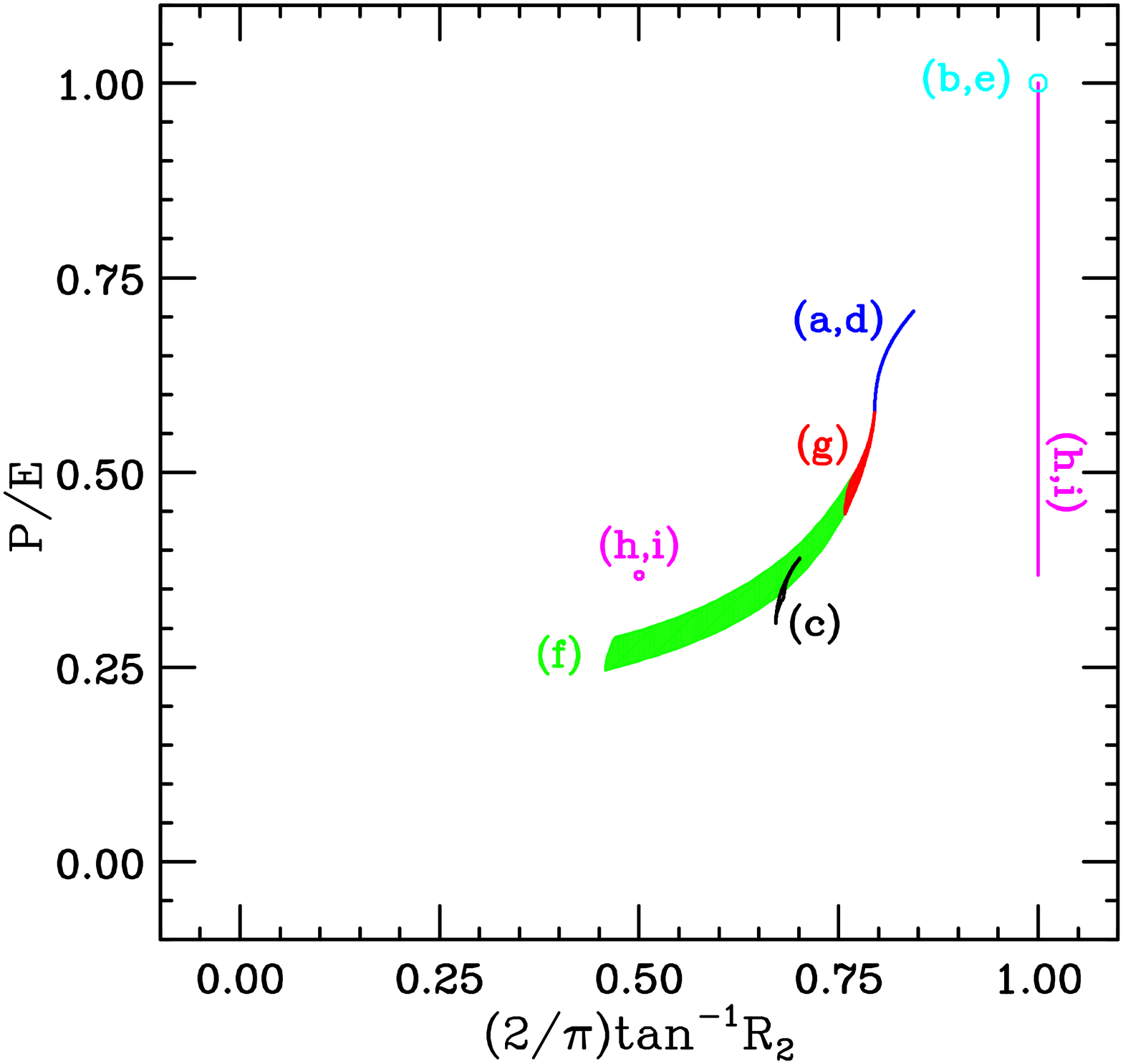}  
\caption{\label{fig:money} The topology disambiguation diagram. The different color-coded regions delineate 
the range of values for $R_2$ and $P/E$ spanned by each decay topology from Fig.~\ref{fig:topologies}. }
\end{figure}

{\it The topology of Fig.~\ref{fig:topologies}(d).}  The invariant mass distribution of the 
visible particles $v_1$ and $v_2$ is not affected by the emission 
of invisible particles upstream and so this case is equivalent 
to the topology of Fig.~\ref{fig:topologies}(a). The corresponding results
can be obtained from (\ref{eq:fa}-\ref{eq:R2a}) with the substitution $A\to B$, since
now the role of the parent resonance is played by the intermediate particle $B$.
One would then be able to determine independently $M_B$ and $M_{\chi_2}$,
while $M_A$ and $M_{\chi_1}$ would remain unknown.

{\it The topology of Fig.~\ref{fig:topologies}(e).} Similarly, this case is equivalent to
Fig.~\ref{fig:topologies}(b), with the substitutions $A\to B_1$, $B\to B_2$ and $\chi\to \chi_2$.
Once again, the emission of the invisible particle $\chi_1$ upstream is not observable.
The only measurable parameter in this case will be the endpoint $E$.

 {\it The topology of Fig.~\ref{fig:topologies}(f).} We find
\bea
f(m)&\sim& m\int^{(M_A-M_{\chi_1})^2}_{M_{B}^{2}(1+\frac{m^2}{M_B^2-M_{\chi_2}^2})}
 \frac{ds}{s}\sqrt{\lambda(s,M_A^2,M_{\chi_1}^2}) ~~~  \nonumber \\
&\sim& -m \left[ K_+K_- + {\frac{1}{2}(X_+^2+X_-^2)} \ln \left(\frac{K_++K_-}{K_+-K_-}\right)  \right.  \nonumber \\
&&  \qquad\quad \left. + {X_+X_-}\ln \left(\frac{X_-K_+-X_+K_-}{X_-K_++X_+K_-}\right) \right], ~~~~
\label{eq:ff}
\eea
where
\bea
X_\pm & \equiv& M_A\pm M_{\chi_1},\quad K_{\pm}\equiv \sqrt{X_\pm^2 -K^2(m)},  ~~~ \\
K^2(m)&\equiv& M_B^2\left(1+\frac{m^2}{M_B^2-M_{\chi_2}^2}\right), \\
E &=&  \sqrt{((M_A-M_{\chi_1})^2-M_B^2)(1-M_{\chi_2}^2/M_B^2)}. ~~~~
\label{eq:Ef}
\eea
In this case, out of the 4 input masses entering the topology of Fig.~\ref{fig:topologies}(f),
one can measure three independent degrees of freedom, e.g.
$M_A/M_B$, $M_{\chi_1}/M_B$ and $M_B^2-M_{\chi_2}^2$.

{\it The topology of Fig.~\ref{fig:topologies}(g).} The shape is described by
\beq
f(m)\sim m\int^{M_{B}^{2}(1-\frac{m^2}{M_A^2-M_B^2})}_{(M_{\chi_1}+M_{\chi_2})^2}
\frac{ds}{s}\sqrt{\lambda(s,M_{\chi_1}^2,M_{\chi_2}^2})
\eeq
and it is easy to see that the results are obtained from (\ref{eq:ff}-\ref{eq:Ef})
with the substitution $M_A\leftrightarrow - M_{\chi_2}$. In particular, the three measurable 
parameters in this case can be taken as $M_{\chi_1}/M_B$, $M_{\chi_2}/M_B$ and $M_A^2-M_B^2$.
 
{\it The topology of Fig.~\ref{fig:topologies}(h).} 
This is the ``sandwich" topology studied in \cite{Agashe:2010gt}.
The shape is given by
\bea
f(m) &\sim& \left\{    
\begin{array}{ll}
\eta\, m,                          & 0\le m\le e^{\minus \eta}\,E, \\ [2mm]
m \ln\left( E/m\right),    & e^{\minus \eta}\,E \le m \le E,
\end{array}
\right.
\label{eq:fh} \\
\eta &\equiv&  
\cosh^{\minus 1} \left(\frac{M^2_{B_1}+M_{B_2}^2-M_{\chi_1}^2}{2M_{B_1}M_{B_2}} \right),
\eea
and 
\bea 
E &=&  \left[e^{\eta} (M_A^2-M_{B_1}^2)(M_{B_2}^2-M_{\chi_2}^2) /(M_{B_1}M_{B_2})\right]^{1/2},~~~~~ \\
P &=& \left\{   
\begin{array}{ll}
E e^{\minus \eta},    &  \eta <1;\\
E e^{\minus 1},         &  \eta \ge 1;
\end{array}
\right. \qquad
R_2 =   \left\{   
\begin{array}{ll}
\infty,    &   \eta <1; \\
1,        &  \eta \ge 1.
\end{array}
\right. 
\label{eq:Ph}
\eea
The distribution (\ref{eq:fh}) exhibits a cusp at the non-differentiable point $m=e^{-\eta}E$. 
In this case, there are 5 mass inputs:
$M_A$, $M_{B_1}$, $M_{B_2}$, $M_{\chi_1}$ and $M_{\chi_2}$, but only two independent measurable
parameters: $\eta$ and $E$.

{\it The topology of Fig.~\ref{fig:topologies}(i).} This is the ``antler" topology which was studied in
\cite{Han:2009ss} for the symmetric case of $M_{B_1}=M_{B_2}$ and $M_{\chi_1}=M_{\chi_2}$. 
Here we generalize the result in \cite{Han:2009ss} to arbitrary masses and find that $f(m)$ is 
given by the same expression (\ref{eq:fh}), only this time
\bea
\eta &\equiv&   \cosh^{\minus 1} 
 \left(\frac{M^2_{A}-M_{B_1}^2-M_{B_2}^2}{2M_{B_1} M_{B_2}} \right), \\
E  &=&   \left[e^{\eta} (M_{B_1}^2-M_{\chi_1}^2)(M_{B_2}^2-M_{\chi_2}^2) /(M_{B_1}M_{B_2})\right]^{1/2}~~~~~~
\eea
and identical expressions (\ref{eq:Ph}) for $P$ and $R_2$.
Just like the case of Fig.~\ref{fig:topologies}(h), out of the 5 mass inputs, 
$\eta$ and $E$ are the only two measurable mass parameters.
Table~\ref{tab:parameters} summarizes the final tally of input particle masses 
and independent measurable parameters for each topology. 

\begin{table}[b]
\caption{\label{tab:parameters}
The number of mass inputs $N_m$ for each topology in Fig.~\ref{fig:topologies}
and the number of independent measurable parameters $N_p$ in the definition of $f(m)$.}
\begin{ruledtabular}
\begin{tabular}{ c | c  c  c  c c}
Topology   &  (a,d)  &  (b,e) &  (c)  &   (f,g)   &   (h,i)         \\   \hline
$N_m$     &     2      &    3    &  3    &    4       & 5           \\ 
$N_p$      &     2      &     1   &  3    &    3       & 2           \\
 \end{tabular}
\end{ruledtabular}
\end{table}


Each topology from Fig.~\ref{fig:topologies} also maps onto a restricted region in 
the $(R_2,P/E)$ plane, as shown in Fig.~\ref{fig:money} (for convenience, instead of $R_2\in (0,\infty)$,
in the figure we plot $\frac{2}{\pi}\tan^{\minus 1}R_2\in (0,1)$). For example,
the cyan circle at $(1,1)$ marks the prediction for the two topologies of Fig.~\ref{fig:topologies}(b,e), 
while the magenta dot at $(0.5,0.37)$ and the magenta vertical line correspond to the two topologies of 
Fig.~\ref{fig:topologies}(h,i). The blue (red, green, black) points refer to the topologies of 
Fig.~\ref{fig:topologies}(a,d) (Fig.~\ref{fig:topologies}(g), Fig.~\ref{fig:topologies}(f), Fig.~\ref{fig:topologies}(c)).
Fig.~\ref{fig:money} demonstrates that with the three measurements $E$, $P$ and $R_2$,
one can already begin to constrain qualitatively the allowed event topologies. 

\begin{figure}[t]
\includegraphics[width=9.5cm]{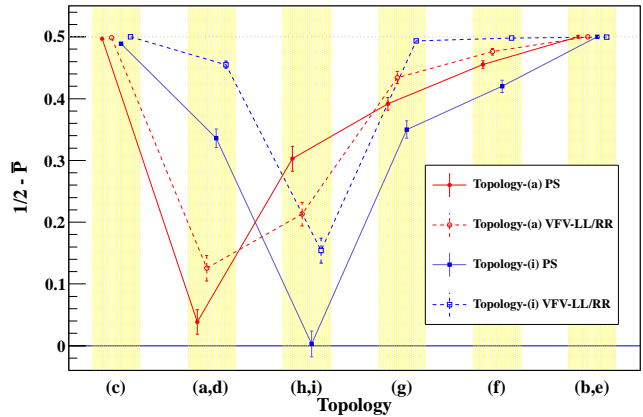}  
\caption{\label{fig:money2} Results from a quantitative topology disambiguation exercise using $\chi^2$ as a test statistics.}
\end{figure}

In fact, one can do even better by fitting to the full invariant mass shapes derived here. 
For illustration, we consider a scenario where particle $A$ is a vector boson (V), 
$B$ is a fermion (F) and $C$ is another vector boson (V), and study two representative 
event topologies. The blue squares in Fig.~\ref{fig:money2} correspond to 
the antler topology case of Fig.~\ref{fig:topologies}(i) for which the $m$ distribution
exhibits a cusp at $m=e^{-\eta}E$ (see Eq.~(\ref{eq:fh})).
This example was considered in \cite{Han:2009ss} 
for the purpose of measuring the masses, which were chosen as
$M_A=1500$ GeV, $M_{B_1}=M_{B_2}=730$ GeV and $M_{\chi_1}=M_{\chi_2}=100$ GeV.
We also consider one cusp-less case, namely, the topology of Fig.~\ref{fig:topologies}(a) 
with a mass spectrum $M_A=550$ GeV, $M_\chi=400$ GeV (red circles in Fig.~\ref{fig:money2}).

Fig.~\ref{fig:money2} shows the average $p$-values ($\bar{P}$) obtained in 200 pseudo-experiments,
with 500 events each. For each example, the filled symbols represent the case in which spin correlations 
are absent, i.e.,~the ``data" is sampled from the phase space distributions derived earlier. 
We see that the fit clearly prefers the correct topologies from Fig.~\ref{fig:topologies}(i) 
and Fig.~\ref{fig:topologies}(a) (and their identical twins from Figs.~\ref{fig:topologies}(h)
and \ref{fig:topologies}(d)), while the wrong topologies are disfavored. 

In models in which the fermions have chiral couplings, the invariant mass shapes considered here
will be slightly distorted due to spin correlations \cite{Burns:2008cp}.
In order to study the effect of spins, we repeat the two exercises for the case of
purely left-handed (L) or purely right-handed (R) couplings of the fermions $B$ 
to the vector bosons $A$ and $C$. The results are displayed in Fig.~\ref{fig:money2} 
with open symbols. We see that, even though we were fitting to pure phase space
formulas\footnote{Following the procedure in \cite{Burns:2008cp}, one could derive analytical formulas for the 
invariant mass distributions in the presence of spin effects, but that is beyond the scope of the current paper \cite{CKMP}.}, 
the correct topologies are still singled out, as they provide the best fit to the data. 

\vspace{-0.5cm} 
\acknowledgments
\noindent
WSC thanks the US National Science Foundation, grant NSF-PHY-0969510, the LHC Theory Initiative.
DK acknowledges support from the LHC Theory Initiative graduate fellowship (NSF Grant No. PHY-0969510).
MP is supported by the CERN-Korea fellowship through National Research Foundation of Korea.
Work supported in part by U.S. Department of Energy Grant DE-FG02-97ER41029

\end{document}